\documentstyle[prl,aps,twocolumn,amssymb,epsfig,graphicx]{revtex}

\newcommand{\gtsim}{\mathop{\,>\kern-1.05em\lower1.ex\hbox{$\sim$}\,}}
\newcommand{\la}{\mathord{\langle}}
\newcommand{\ra}{\mathord{\rangle}}

\begin{document}
\draft

\title{Theory for the enhanced induced magnetization in coupled
  magnetic trilayers in the presence of spin fluctuations }

\author{P.J. Jensen,$^{1,2}$ K.H. Bennemann,$^2$ }

\address{
$^1$ Hahn-- Meitner-- Institut, Glienicker Str.100, 
D -- 14 109 Berlin, Germany} 

\address{$^2$ Institut f\"ur Theoretische Physik, Freie Universit\"at 
Berlin \\ Arnimallee 14, D -- 14 195 Berlin, Germany} 

\author{P. Poulopoulos,  M. Farle,$^*$ F. Wilhelm, K. Baberschke}

\address{Institut f\"ur Experimentalphysik, Freie Universit\"at 
Berlin \\ Arnimallee 14, D -- 14 195 Berlin, Germany} 

\date{\today}
\maketitle

\begin{abstract}
Motivated by recent experiments, 
the effect of the interlayer exchange interaction $J_{inter}$ on the  
magnetic properties of coupled Co/Cu/Ni trilayers is studied theoretically. 
Here the Ni film has a lower Curie temperature $T_{C,\rm Ni}$ than the 
Co film 
in case of decoupled layers. We show that by taking into account magnetic 
fluctuations the interlayer coupling induces a strong magnetization for 
$T\gtsim T_{C,\rm Ni}$ in the Ni film. For an increasing $J_{inter}$ 
the resonance-like peak of the longitudinal Ni susceptibility 
is shifted to larger temperatures, whereas its maximum value decreases 
strongly. A decreasing Ni film thickness enhances the induced Ni 
magnetization for $T\gtsim T_{C,\rm Ni}$. The measurements cannot be 
explained properly by a mean field estimate, which yields a ten 
times smaller effect. Thus, the observed magnetic properties indicate 
the strong effect of 2D magnetic fluctuations 
in these layered magnetic systems. The calculations are performed with 
the help of a Heisenberg Hamiltonian and a Green's function approach. 
\end{abstract}

Recently, the element specific magnetization and the longitudinal 
susceptibility of magnetic epitaxial Co/Cu/Ni trilayers grown on Cu(001) 
have been measured \cite{BWS98,NWF99}. 
The two ferromagnetic Ni and Co films are coupled by the 
indirect exchange interaction $J_{inter}$ across the nonmagnetic Cu
layer, which exhibits an oscillatory behavior as a function of 
the thickness $d_{\rm Cu}$ of the spacer, and an overall decay like 
$d_{\rm Cu}^{-2}$ \cite{NWF99,BrC91,BBK96}. 
The thicknesses $d_{\rm Ni}$ and $d_{\rm Co}$ of the Ni and Co films 
are chosen in such a way that for a vanishing interlayer coupling 
the Ni film has a lower Curie temperature than the Co film, i.e.\ 
$T_{C,\rm Ni}(d_{\rm Ni})<T_{C,\rm Co}(d_{\rm Co})$. $J_{inter}$ 
induces for $T>T_{C,\rm Ni}$ 
a considerable magnetization in the Ni film, which has been measured to 
vanish $\sim 30 - 40$ K above $T_{C,\rm Ni}$. In this work we show 
that the induced strong Ni magnetization can theoretically only be obtained 
properly by taking into account magnetic fluctuations in the Ni film. 
If these fluctuations are neglected in the calculations 
(for example within a mean field theory (MFT) approach \cite{WaM92}), 
the resulting induced Ni magnetization $M_{\rm Ni}(T)$ for 
$T\gtsim T_{C,\rm Ni}$ is an order of magnitude smaller. Vice versa, 
the neglect of these fluctuations requires an unrealistic large 
value for $J_{inter}$ to yield the observed shift $\Delta T$ of 
$M_{\rm Ni}(T)$ to larger temperatures. Generally, spin 
fluctuations diminish the magnetization of a two-dimensional (2D) 
magnetic system more strongly than for bulk magnets \cite{Mal76,Yab91}. 
An external magnetic field suppresses the action of these fluctuations, 
resulting in a stronger increase of the magnetization in 2D than in 
bulk systems. Similarly, in case of a coupled trilayer the interlayer 
coupling reduces the fluctuation effect, since it acts as an external 
magnetic field. Consequently, $J_{inter}$ has a pronounced effect on the 
Ni film magnetization. The magnetic behavior of such a 
system can be used to study the action of the strong 2D spin fluctuations.  

To take into account the collective magnetic excitations (spin waves, 
magnons), we apply a many-body Green's function approach,  
and use the so-called Tyablikov (or RPA-) decoupling \cite{Tya59}. 
Since within this method interactions between magnons are partly 
taken into account, 
the whole temperature range of interest up to the Curie temperature 
can be considered.  A Heisenberg-type Hamiltonian on an
fcc(001) thin film with $d=d_{\rm Ni}+d_{\rm Co}$ monolayers (ML) is 
assumed with localized magnetic moments 
$\mbox{\boldmath$\mu$}_i=\mu_i\,{\bf S}_i/S$ on lattice sites $i\;$: 
\begin{eqnarray}
{\cal H}&=&-\frac{1}{2}\sum_{\langle i,j\rangle}J_{ij}\,{\bf S}_i\,{\bf S}_j 
-\sum_i{\bf B}\,\mbox{\boldmath$\mu$}_i \nonumber \\ &&
+\frac{1}{2}\sum_{i,j \atop i\ne j}\frac{1}{r^5} \Big[
\mbox{\boldmath$\mu$}_i\,\mbox{\boldmath$\mu$}_j\,r^2
-3({\bf r}\,\mbox{\boldmath$\mu$}_i )
({\bf r}\,\mbox{\boldmath$\mu$}_j) \Big]\,. \label{e1} 
\end{eqnarray}  
Quantum mechanical spins with spin quantum number $S=1$ are assumed. 
Due to the dipole interaction the magnetization $M_i(T)=\la S_i^z\ra$ 
is directed in-plane, determining the 
quantization axis ($z$-direction). The external magnetic field 
${\bf B}=(0,0,B)$ is applied parallel to this axis. 
$J_{ij}$ are the exchange couplings between nearest neighbor spin 
pairs which are chosen in such a way that they yield the observed Curie 
temperatures for the separate (i.e.\ decoupled) layers. We put 
$J_{\rm CoCo}=398$ K per bond to obtain $T_{C,\rm Co}(2)=435$ K for a Co film 
with $d_{\rm Co}=2$ ML \cite{BWS98}. To account for the diminished interface 
magnetic state of the Ni film, we distinguish between exchange 
couplings in the interface layers and the film interior layers. With 
$J_{\rm NiNi}^{interface}=30$ K and $J_{\rm NiNi}^{interior}=172$ K per bond 
one obtains for a Ni film with $d_{\rm Ni}=5$ ML $T_{C,\rm Ni}(5)=267$ K 
\cite{NWF99}. These numbers for the exchange couplings are somewhat lower 
than the corresponding values obtained from the bulk Curie temperatures.  
In addition an interlayer exchange coupling $J_{inter}$ across the
nonmagnetic Cu spacer layer between Ni and Co spins in the interlayer 
next to Cu is assumed. Positive as well as negative values of 
$J_{inter}$ can be considered, preferring thus parallel ($J_{inter}>0$) or 
antiparallel ($J_{inter}<0$) magnetized Ni and Co layers. 
The last term in Eq.(\ref{e1}) is the magnetic dipole coupling between
spins $\mbox{\boldmath$\mu$}_i$ and $\mbox{\boldmath$\mu$}_j$ 
separated by vectors ${\bf r}={\bf r}_j-{\bf r}_i$, denoting $r=|{\bf r}|$.   
The slowly converging oscillating lattice sums are converted into 
rapidly converging ones with the help of Ewald summation \cite{BeM69}. 
Layer-dependent magnetic moments $\mu_i$ are assumed \cite{SWN98}. In
particular we put $\mu_{\rm Ni}^{interface}=0.46\;\mu_B$, 
$\mu_{\rm Ni}^{interior}=0.61\;\mu_B$, and $\mu_{\rm Co}=2.02\;\mu_B$ 
for all Co layers, $\mu_B$ is the Bohr magneton. 
Lattice anisotropy terms are not considered here. 

For the calculation of the layer-dependent magnetizations, 
$M_i(T)$, $i=1\ldots d$, we consider the following 
two-times (commutator-) Green's functions, which  
are written in spectral representation as \cite{Tya67,Cal63} 
\begin{equation}
G_{ij}^{+-(n)}(\omega,{\bf k_\|})=
\la\la S_i^+; (S_j^z)^n\,S_j^-\ra\ra_{\omega,{\bf k_\|}}=
\la\la S_i^+; C_j^{(n)}\ra\ra_{\omega,{\bf k_\|}} \,. \label{e3} 
\end{equation}
Here the $i,j$ refer to layer indices. Because we assume ferromagnetically 
ordered layers, the lateral periodicity has been used to apply a Fourier 
transformation into the 2D momentum space, {\bf k}$_\|\ $ being the 2D wave 
vector. The Green's functions are determined by solving the familiar 
equation of motion. Higher order Green's functions are approximated by 
the Tyablikov (RPA-) decoupling \cite{Tya59} for the exchange and 
dipole interactions ($i\neq k$): 
\begin{equation}
\la\la S_i^z\,S_k^+;C_j^{(n)}\ra\ra \approx 
\la S_i^z\ra\,\la\la S_k^+;C_j^{(n)}\ra\ra = 
M_i(T)\,G_{kj}^{+-(n)} \,, \label{e4} 
\end{equation}
i.e.\ spin operators $S^z_i$ are replaced by their expectation values 
$M_i(T)$. Different integers $n\le 2S-1$ have to be
considered in order to calculate different spin quantum numbers
$S$ \cite{Tya67}. The equations of motion lead to a set of $d$ coupled
linear equations for the Green's functions. With the help of the 
spectral theorem the respective expectation values (or correlation functions) 
$\la(S_j^z)^nS_j^-S_i^+\ra$ are determined. The magnetization $M_i(T)$
is obtained from the usual relations between spin operators. 
By comparison with a recent Quantum Monte Carlo 
calculation of a Heisenberg monolayer in an external magnetic field 
\cite{TGH98} it was shown that the applied Green's function method yields 
a satisfactory result for the magnetization \cite{EFJ99,HoP40}. 

\begin{figure}
\includegraphics[width=6cm,height=8cm,angle=-90,bb=90 40 540 750
,clip]{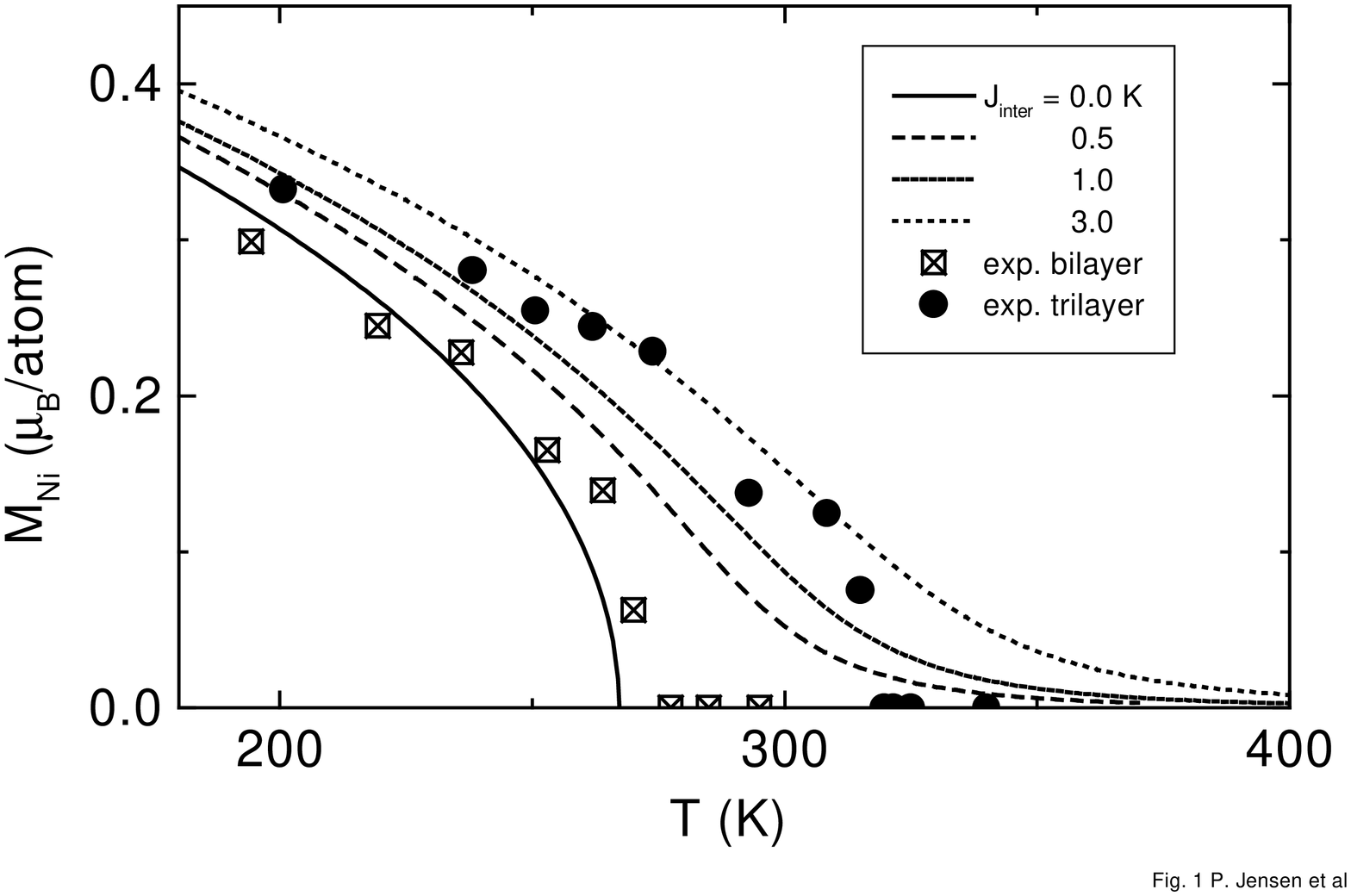}
\vspace{0.3cm}

\caption{Ni magnetization $M_{\rm Ni}(T)$ for a Co/Cu/Ni 
trilayer as a function of the temperature $T$ calculated by the Green's
function approach. Different interlayer couplings $J_{inter}$
(in K/bond) are assumed as indicated. An epitaxial trilayer with
$d_{\rm Ni}=5$ Ni and $d_{\rm Co}=3$ Co monolayers (ML) is assumed. 
The values for the exchange 
couplings within the Ni and Co films and the corresponding magnetic
moments are given in the text. $T_{C,\rm Ni}=$ 267 K is the Curie 
temperature of the single Ni film. In addition 
experimental results for the Ni magnetization are displayed for a 
single 4.8 ML Ni film capped with Cu ('bilayer'), as well as for a 
Co/Cu/Ni trilayer with $d_{\rm Ni}=4.8$ ML, $d_{\rm Co}=2.8$ ML, 
and $d_{\rm Cu}=2.8$ ML [2]. }  
\end{figure}
In Fig.(1) we present results for the Ni magnetization $M_{\rm Ni}(T)$ 
as a function of the temperature $T$ calculated with 
different interlayer couplings $J_{inter}$. We consider a Co/Cu/Ni trilayer 
with $d_{\rm Ni}=5$ Ni, $d_{\rm Cu}=3$ Cu, and $d_{\rm Co}=3$ Co monolayers, 
respectively. For comparison experimental results for the same system are 
also shown \cite{NWF99}. The layer-dependent magnetizations 
$M_i(T)$, $i=1\ldots d$, are determined from an iterative procedure. 
Presented are the Ni magnetizations $M_{\rm Ni}(T)$ averaged over all 
Ni layers. We use the inflection point $T_{infl}$ of $M_{\rm Ni}(T)$
as a measure of the corresponding temperature shift 
$\Delta T(J_{inter})=T_{infl}-T_{C,\rm Ni}$ of the Ni magnetization 
with respect to the decoupled case. One observes that already a small 
value of the interlayer coupling $J_{inter}$ produces a comparably 
large $\Delta T(J_{inter})$. For example, $J_{inter}=1$ K results in 
$\Delta T\approx 30$ K. Such a value for $J_{inter}$ 
compares well with various results measured formerly with different 
methods \cite{BBK96,Par91}. Corresponding results have been  
determined by us also from a MFT approach. For the same value of 
$J_{inter}$ the calculated $\Delta T(J_{inter})$ 
obtained from this approximation is about 10 times smaller 
than the value resulting from the Green's function method. 

We stress that this strong difference is a result of the 2D character 
of the magnetic trilayer system. The action of an external magnetic 
field for $T\gtsim T_C$ is much more pronounced for a 2D magnet than for 
a corresponding bulk system \cite{Yab91}. 
For the coupled trilayer system under consideration the interlayer coupling 
$J_{inter}$ acts similar as an external magnetic field. Therefore, 
for temperatures $T\gtsim T_{C,\rm Ni}$ close to the Curie
temperature of the single Ni film already a small $J_{inter}$ is sufficient
to induce a marked Ni magnetization and the corresponding temperature 
shift $\Delta T$. In contrast, within a MFT approach the exchange coupling
alone results in a finite remanent magnetization for a 2D magnet, and 
does not need the support of the dipole coupling or an external magnetic 
field. In this case a small interlayer coupling adds simply to the strong 
Ni-Ni exchange coupling and results in a correspondingly small value of 
$\Delta T$. 

\begin{figure}
\hspace*{-0.5cm}\includegraphics[width=9cm,height=10.5cm,bb=35 90 540 690,clip]
{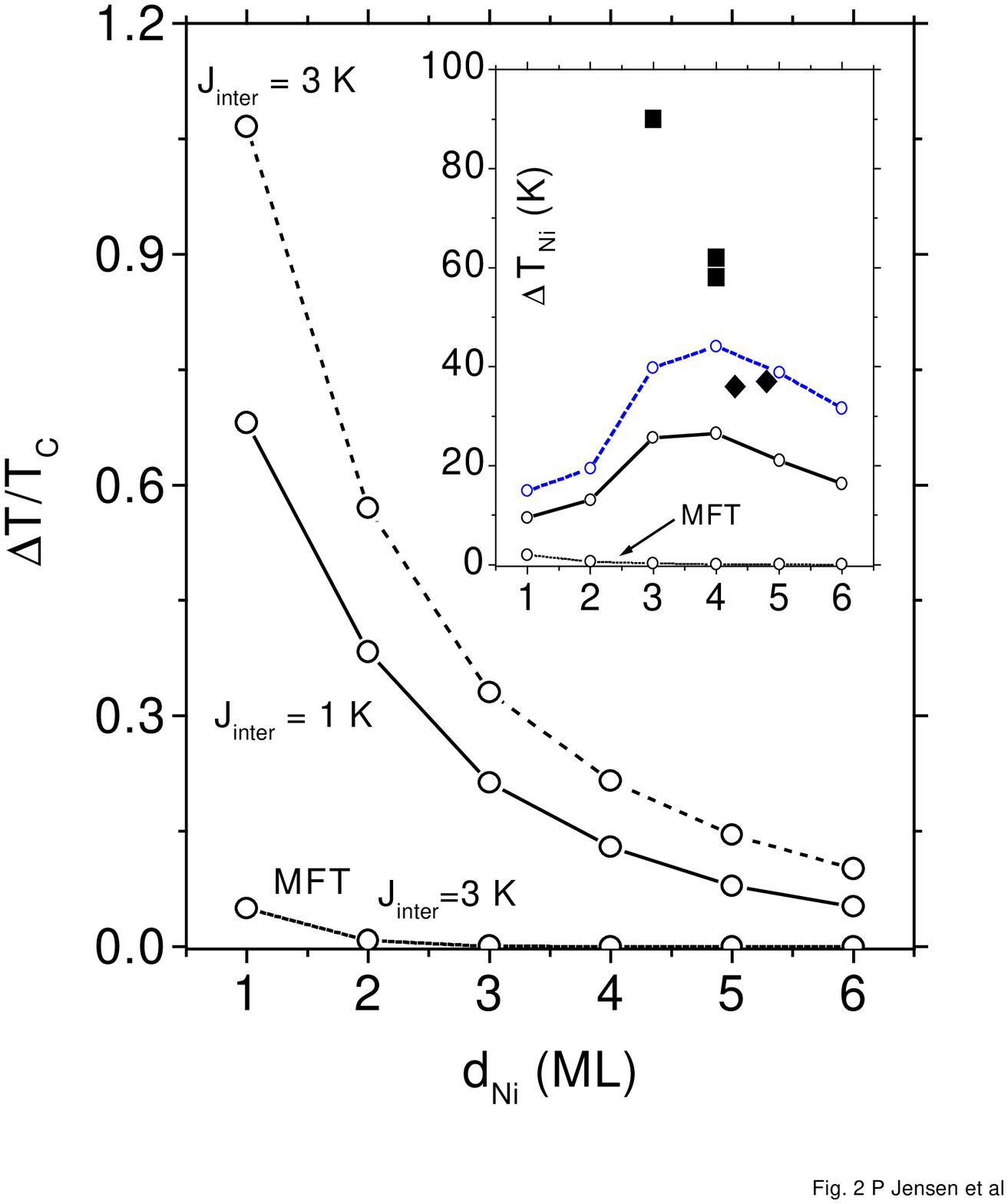}

\caption{
Temperature difference $\Delta T(d_{\rm Ni})$ between the inflection 
point of the Ni magnetization $M_{\rm Ni}(T)$ in the coupled Co/Cu/Ni trilayer 
and the Curie temperature $T_{C,\rm Ni}(d_{\rm Ni})$ of the single Ni 
film as a function of the Ni film thickness $d_{\rm Ni}$. Two
different interlayer couplings $J_{inter}=$ 1 K and 3 K are assumed,
the other system parameters are the same as for Fig.1. The figure 
displays the relative temperature difference $\Delta T$ scaled by 
$T_{C,\rm Ni}(d_{\rm Ni})$. In the inset the absolute temperature shift 
is given. For comparison we show the respective results for $\Delta T$ 
calculated by an MFT approach, assuming $J_{inter}=3$ K. 
Experimental results are also given for two different Cu
thicknesses: $\blacklozenge$: $d_{\rm Cu}=2.8$ ML; and $\blacksquare$: 
$d_{\rm Cu}=2.0$ ML, the latter corresponds to a larger $J_{inter}>3$ K.}
\end{figure}

We have tested the assumption that the interlayer coupling acts similar 
as an external magnetic field. The results of $M_{\rm Ni}(T)$ for a 
single (decoupled) Ni film with $d_{\rm Ni}=5$ ML, with an external 
magnetic field acting exclusively on the topmost Ni layer with a strength 
$B=J_{inter}/\mu_{\rm Ni}$, are practically 
the same as for the corresponding coupled trilayer system. 

Furthermore, we have calculated the induced Ni magnetization at 
$T\gtsim T_{C,\rm Ni}$ for different thicknesses $d_{\rm Ni}$ of the Ni film. 
Results for $\Delta T(d_{\rm Ni})$ are shown in Fig.(2) for two different 
values of $J_{inter}$ and for $1\le d_{\rm Ni}\le 6$ ML. The other coupling 
parameters are the same. The resulting \em absolute \em value 
$\Delta T(d_{\rm Ni})$ exhibits a maximum at about $d_{\rm Ni}=4$ ML,
see inset of Fig.(2). For comparison also the corresponding results 
calculated by the MFT approach are shown. On the other hand, the 
\em relative \em temperature shift $\Delta T/T_{C,\rm Ni}(d_{\rm Ni})$, 
scaled by the Curie temperature $T_{C,\rm Ni}(d_{\rm Ni})$ of the single 
Ni film, increases by reducing the thickness of the Ni film, s.\ Fig.(2). 
This indicates the increasing importance of the action of the magnetic 
fluctuations for a decreasing film thickness. Experimental results are 
also displayed for two different Cu spacer thicknesses. 

\begin{figure}
\hspace*{-0.5cm}\includegraphics[width=6cm,height=9cm,angle=-90,
bb=70 40 550 750,clip]{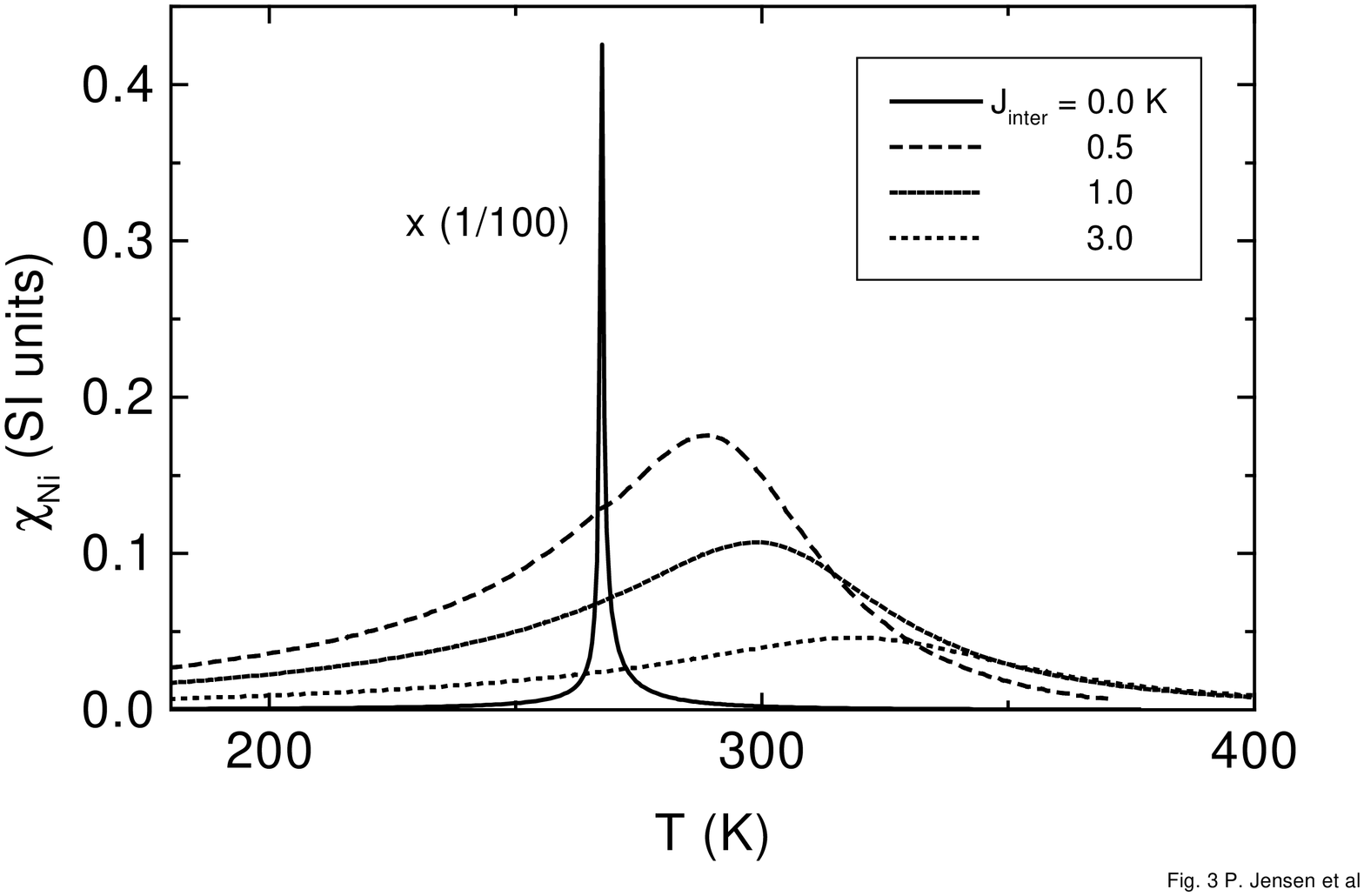}
\vspace{0.3cm}

\caption{
Ni film susceptibility $\chi_{\rm Ni}(T)$ as a function of the 
temperature for different interlayer couplings as indicated. The 
susceptibility field amplitude is 2 G. The same Co/Cu/Ni trilayer 
system as described in Fig.(1) is assumed. Note that the susceptibility 
curve for $J_{inter}=0$ is scaled by the factor 1/100. }
\end{figure}

In addition we investigate the longitudinal susceptibility $\chi_{\rm Ni}(T)$ 
of the Ni film. For this purpose $\chi_{i}(T)$ for the $i$th magnetic layer,
$i=1\ldots d$, is calculated from the difference of the magnetizations 
$M_i(T,B)$ with $B=0$ and $B=2$ G,  
\begin{equation}
\chi_i^{zz}(T)\equiv\chi_i(T)=\frac{\partial M_i(T,B)}{\partial B}
\approx\frac{\Delta M_i(T,B)}{\Delta B} \,. \label{e2} 
\end{equation}
In Fig.(3) the Ni susceptibility $\chi_{\rm Ni}(T)$ averaged over all Ni 
layers is displayed, corresponding to the results of Fig.(1). A 
resonance-like peak of $\chi_{\rm Ni}(T)$ is obtained for $T\gtsim
T_{C,\rm Ni}$ as has been reported previously \cite{WaM92}. With 
increasing interlayer coupling $J_{inter}$ the susceptibility peak is 
shifted to higher temperatures. 
Also, the maximum value of $\chi_{\rm Ni}(T)$ reduces strongly and its 
corresponding width increases markedly \cite{WaM92}. For a strong 
$J_{inter}$ the Ni susceptibility is so small that it may be hardly 
measurable. A singularity of $\chi_{\rm Ni}(T)$ will occur at
$T=T_{C,\rm Co}$ since there the induced Ni magnetization, although 
small, vanishes in accordance with the vanishing Co magnetization. 
Thus, $T_{C,\rm Co}$ corresponds to the true phase transition 
temperature of the coupled magnetic Co/Cu/Ni trilayer system \cite{WaM92}.   

In Refs.\ \cite{BWS98,NWF99} it was found for the Co/Cu/Ni trilayer system 
that the observed Ni remanent magnetization \em vanishes \em above a
temperature $T_{\rm Ni}^*$, where $T_{C,\rm Ni}<T_{\rm Ni}^*<T_{C,\rm Co}$. 
On the other hand, either no 
susceptibility signal or only a small peak in $\chi_{\rm Ni}(T)$ could be 
measured at $T_{\rm Ni}^*$. This might be due e.g.\ to the occurrence of a 
multidomain state or to a magnetic 
reorientation in the Ni film. A true phase transition in the 
thermodynamic sense is reminiscent to a nonanalytical behavior of the 
free energy, resulting in singularities of e.g.\ the correlation length 
or the susceptibility, as found at $T_{C,\rm Co}$. Regardless of the 
particular behavior of the magnetic properties at $T_{\rm Ni}^*$, we 
emphasize that the observed strong induced Ni magnetization and the shift 
of the Ni susceptibility for $T\gtsim T_{C,\rm Ni}$ due to the interlayer 
coupling $J_{inter}$ is caused by the presence of magnetic fluctuations. 
To compare the measured and calculated temperature shift 
$\Delta T$ we have assumed $T_{infl}\approx T_{\rm Ni}^*$. 

We have investigated asymmetric trilayers which
exhibit different Curie temperatures for the decoupled Ni and Co films, 
e.g.\ for single magnetic layers or a thick nonmagnetic spacer.  
The interlayer coupling $J_{inter}$ influences mainly the
magnetization of the Ni film, which has the lower ordering temperature,  
whereas $T_{C,\rm Co}$ stays practically constant. On the other hand, 
a symmetric system, e.g.\ a Ni/Cu/Ni trilayer with equal Ni film
thicknesses or a periodic multilayer system, has a single Curie 
temperature $T_C$. Here $J_{inter}$ will enhance $T_C$ considerably 
by amounts similar as discussed for the asymmetric trilayer. Indeed 
this has been observed for a Ni/Au multilayer system \cite{BBK96}. 
In principle, by an appropriate combination of tri- and multilayers, 
for instance by varying the materials and the thicknesses of the 
magnetic and nonmagnetic layers, one might tune the magnetic properties 
according to possible applications.  

In summary, we have calculated the action of the interlayer exchange 
coupling in a magnetic Co/Cu/Ni trilayer system by means of a
many-body Green's function approach for a Heisenberg Hamiltonian. 
$J_{inter}$ induces a considerable magnetization in the Ni film 
for $T\gtsim T_{C,\rm Ni}$, and shifts the Ni susceptibility peak  
to larger temperatures. Also the width of the Ni 
susceptibility increases, whereas its maximum value decreases strongly. 
We have shown that for reasonable values of $J_{inter}$ the observed 
strong induced Ni magnetization can be obtained 
only if the magnetic fluctuations in these 2D systems are taken into 
account properly. Corresponding results as calculated by a MFT approach, 
which neglect these fluctuations, yield a 10 times smaller increase, 
and cannot explain the observed magnetic behavior for the Co/Cu/Ni 
trilayer system. The influence of the magnetic 
fluctuations becomes stronger for smaller Ni film thicknesses, s.\ 
Fig.(2), indicating the 2D character of the important correlations. 
Note that we have investigated the effect of $J_{inter}$ on the 
magnetic properties solely by considering thermal fluctuations. 
Other possible influences such as magnetic noncollinearities are not 
discussed here. 
  
This work has been supported by the DFG, Sonderforschungsbereich 290. 
Discussions with C.\ Timm are gratefully acknowledged.


\begin{references} 
\vspace*{-1.5cm}
\bibitem[*]{farle} {\it new address:} Institut f\"ur Halbleiterphysik 
und Optik, Technische Universit\"at Braunschweig, Mendelssohnstr. 3, 
D-38 106 Braunschweig, Germany. 
\bibitem{BWS98} U. Bovensiepen, F. Wilhelm, P. Srivastava, P. Poulopoulos, 
M. Farle, A. Ney, and K. Baberschke, Phys. Rev. Lett. {\bf81}, 2368 (1998). 
\bibitem{NWF99} A. Ney, F. Wilhelm, M. Farle, P. Poulopoulos, 
P. Srivastava, and K. Baberschke, Phys. Rev. B {\bf59}, R3938 (1999). 
\bibitem{BrC91} P. Bruno and C. Chappert, Phys. Rev. Lett. {\bf67}, 
1602 (1991). 
\bibitem{BBK96} G. Bayreuther, F. Bensch, and V. Kottler, J. Appl. Phys. 
{\bf79}, 4509 (1996). 
\bibitem{WaM92} R.W. Wang and D.L. Mills, Phys. Rev. B {\bf46}, 
11 681 (1992). 
\bibitem{Mal76} S.V. Maleev, Sov. Phys. JETP {\bf43}, 1240 (1976); 
V.L. Pokrovsky and M.V. Feigel'man, \em ibid., \em {\bf 45}, 291 (1977).
\bibitem{Yab91} D.A. Yablonsky, Phys. Rev. B {\bf44}, 4467 (1991); 
D. Kerkmann, D. Pescia, and R. Allenspach, Phys. Rev. Lett. {\bf68}, 
686 (1992).
\bibitem{Tya59} S.V. Tyablikov, Ukr. Mat. Zh. {\bf11}, 287 (1959).
\bibitem{BeM69} H. Benson and D.L. Mills, Phys. Rev. {\bf178}, 839 (1969);
P.J. Jensen, Ann. Physik {\bf6}, 317 (1997). 
\bibitem{SWN98} P. Srivastava, F. Wilhelm, A. Ney, M. Farle, H. Wende, 
N. Haack, G. Ceballos, and K. Baberschke, Phys. Rev. B {\bf58}, 5701 (1998). 
\bibitem{Tya67} S.V. Tyablikov, \em Methods in the quantum
theory of magnetism, \em (Plenum Press, New York, 1967).
\bibitem{Cal63} H.B. Callen, Phys. Rev. {\bf130}, 890 (1963).
\bibitem{TGH98} C. Timm, S.M. Girvin, P. Henelius, and A.W. Sandvik,
Phys. Rev. B {\bf58}, 1464 (1998).
\bibitem{EFJ99} A. Ecker, P. Fr\"obrich, P.J. Jensen, and P.J. Kuntz,
J. Phys.: Condens. Matter {\bf11}, 1557 (1999).
\bibitem{HoP40} In presence of the dipole coupling also the Green's functions 
$G_{ij}^{--(n)}$ should be taken into account in order to consider
elliptical spin waves [e.g.\ T. Holstein and H. Primakoff, Phys. Rev. 
{\bf58}, 1098 (1940)]. The $G_{ij}^{--(n)}$ are neglected here because the
dipole coupling is small as compared to the exchange coupling for the
systems under consideration. We found that the results for the
magnetic properties are not affected markedly by
considering the $G_{ij}^{--(n)}$, for instance $T_C$ is reduced by 
$\sim 10 \%$. However, the computational effort enhances drastically. 
\bibitem{Par91} S.S.P. Parkin, Phys. Rev. Lett. {\bf67}, 3598 (1991).
\end{references}
\end{document}